\documentclass{pasa}%

\usepackage{graphicx}

\title[GREENBURST commensal fast radio burst search]{GREENBURST: a commensal fast radio burst search back-end for the Green Bank Telescope}

\author[Surnis et al.]{Mayuresh P. Surnis$^{1,2}$\thanks{E-mail: mayuresh.surnis@mail.wvu.edu}, D. Agarwal$^{1,2}$, D.~R. Lorimer$^{1,2}$, X.~Pei$^{3}$, G.~Foster$^{4}$, A.~Karastergiou$^{4,5,6}$, G.~Golpayegani$^{1,2}$, R.~J.~Maddalena$^{7}$, S.~White$^{7}$, W.~Armour$^{8}$, J.~Cobb$^{9}$, M.A.~McLaughlin$^{1,2}$, D.H.E.~MacMahon$^{9}$, A.P.V.~Siemion$^{9,10,11}$, D.~Werthimer$^{9}$ and C.J.~Williams$^{4}$\\
\affil{$^{1}$West Virginia University, Department of Physics and Astronomy, P. O. Box 6315, Morgantown, WV, USA}
\affil{$^{2}$Center for Gravitational Waves and Cosmology, West Virginia University, Chestnut Ridge Research Building,\\ Morgantown, WV, USA}
\affil{$^{3}$Xinjiang Astronomical Observatory, Chinese Academy of Sciences, Urumqi, Xinjiang 830011, China}
\affil{$^{4}$University of Oxford, Sub-Department of Astrophysics, Denys Wilkinson Building, Keble Road, Oxford, OX1 3RH,\\ United Kingdom}
\affil{$^{5}$Physics Department, University of the Western Cape, Cape Town 7535, South Africa}
\affil{$^{6}$Department of Physics and Electronics, Rhodes University, PO Box 94, Grahamstown 6140, South Africa}
\affil{$^{7}$Green Bank Observatory, P.O. Box 2, Green Bank, WV 24944, USA}
\affil{$^{8}$OeRC, Department of Engineering Science, University of Oxford, Keble Road, Oxford OX1 3QG, UK}
\affil{$^{9}$Department of Astronomy, University of California, Berkeley, 501 Campbell Hall \#3411, Berkeley, CA 94720, USA}
\affil{$^{10}$Radboud University, Nijmegen, 6525 HP, the Netherlands}
\affil{$^{11}$SETI Institute, Mountain View, CA 94043, USA}
}%

\jid{PASA}
\doi{10.1017/pas.\the\year.xxx}
\jyear{\the\year}

\usepackage{aas_macros}
\usepackage{hyperref} 
\hypersetup{colorlinks,citecolor=blue,linkcolor=blue,urlcolor=blue}
\usepackage{lscape}
\usepackage{amsmath}	
\usepackage{amssymb}	
\usepackage[dvipsnames]{xcolor}
\usepackage[xindy]{glossaries}
\glsdisablehyper

\newcommand{\pcc}{pc\,cm$^{-3}$}	
\newcommand{\sups}[1]{\textsuperscript{#1}}

\newcommand{\GB}{GREENBURST}

\newacronym{adc}{ADC}{Analog-to-Digital Converter}
\newacronym{askap}{ASKAP}{Australian Square Kilometre Array Pathfinder}
\newacronym{chime}{CHIME}{Canadian Hydrogen Intensity Mapping Experiment}
\newacronym{dcr}{DCR}{digital continuum receiver}
\newacronym{dsp}{DSP}{digital signal processing} 
\newacronym{dm}{DM}{Dispersion Measure}
\newacronym{fov}{FoV}{field of view}
\newacronym{fpga}{FPGA}{Field Programmable Gate Array}
\newacronym{frb}{FRB}{Fast Radio Burst}
\newacronym{fwhm}{FWHM}{Full-Width at Half-Maximum}
\newacronym{gbt}{GBT}{Green Bank Telescope}
\newacronym{gbncc}{GBNCC}{Green Bank Northern Celestial Cap}
\newacronym{gpu}{GPU}{graphics processing unit}
\newacronym{if}{IF}{intermediate frequency}
\newacronym{igm}{IGM}{Intergalactic Medium}
\newacronym{ism}{ISM}{Interstellar Medium}
\newacronym{lna}{LNA}{Low-Noise Amplifier}
\newacronym{pfb}{PFB}{polyphase filterbank}
\newacronym{rfi}{RFI}{Radio-frequency Interference}
\newacronym{rf}{RF}{Radio Frequency}
\newacronym{sefd}{SEFD}{System Equivalent Flux Density}
\newacronym{seti}{SETI}{Search for Extraterrestrial Intelligence}
\newacronym{snr}{S/N}{Signal-to-Noise Ratio}
\newacronym{sps}{SPS}{Single Pulse Search}
\newacronym{udp}{UDP}{User Datagram Protocol}

\hypersetup{draft}

\begin{document}

\begin{frontmatter}
\maketitle

\begin{abstract}
We describe the design and deployment of \GB, a commensal Fast Radio Burst (FRB) search system at the Green Bank Telescope. \GB~uses the dedicated L-band receiver tap to search over the 960--1920~MHz frequency range for pulses with dispersion measures out to $10^4$~\pcc. Due to its unique design, \GB~ will obtain data even when the L-band receiver is not being used for scheduled observing. This makes it a sensitive single pixel detector capable of reaching deeper in the radio sky. While single pulses from Galactic pulsars and rotating radio transients will be detectable in our observations, and will form part of the database we archive, the primary goal is to detect and study FRBs. Based on recent determinations of the all-sky rate, we predict that the system will detect approximately one FRB for every 2$-$3 months of continuous operation. The high sensitivity of \GB~ means that it will also be able to probe the slope of the FRB source function, which is currently uncertain in this observing band.
\end{abstract}

\end{frontmatter}

\section{INTRODUCTION}
\label{sec:intro}

\glspl{frb} are characterized by their millisecond duration and radio-frequency dispersion that far exceeds that predicted to result from interactions with free electrons in the Milky Way \citep{lbm+07,tsb+13}.  Over the past decade, a number of significant observations have been made, including detections from 400~MHz \citep{abb+19a} to 8~GHz \citep{gsp+18,zgf+18}, \glspl{frb} with extreme implied distances \citep{bkb+18}, and the repeating sources FRB~121102 \citep{ssh+16} and FRB~180814 \citep{abb+19b}. The source FRB~121102 has been associated with a dwarf galaxy at redshift $z=0.19$ \citep{mph+17} and shows extreme Faraday rotation  \citep{msh+18}, demonstrating that it is situated in a dense region in its host galaxy.

FRBCAT\footnote{http://frbcat.org} \citep{pbj+16} provides an up-to-date catalog of reported \gls{frb} detections with 65 reported \glspl{frb} at the time of writing this paper. Recently, there has been a significant increase in the number of reported \glspl{frb} as new wide \gls{fov} arrays have began operating. The \gls{askap}, when observing in `fly's eye' mode to maximize sky coverage, has detected 20 \glspl{frb} \citep{smb+18}. Initial observations with \gls{chime} have also resulted in 13 \gls{frb} detections \citep{abb+19a} including the repeating source FRB~180814 \citep{abb+19b}.

Prior to \gls{askap} and \gls{chime}, most of the \glspl{frb} were detected with the 64-m Parkes radio telescope. Only two \glspl{frb} have been detected with the Robert C.~Byrd  \gls{gbt}. The first, FRB~110523 \citep{mls+15}, was detected in archival \gls{gbt} data acquired with the the prime focus 800-MHz receiver (700--900~MHz). This detection -- with a telescope other than Parkes -- provided strong evidence that \glspl{frb} were in fact astrophysical. Recently, multiple bursts from the repeating source FRB\,121102 were detected \citep{gsp+18,zgf+18} using the \gls{gbt} C-Band (4$-$8~GHz) receiver, while the \gls{gbncc} Pulsar Survey at 350~MHz reported a non-detection of \glspl{frb} \citep{ckj+17}. Recently, \cite{gl19} also reported non-detection of \glspl{frb} at L-band using the 20 m telescope on the GBT site.

Wide \gls{fov} arrays have been very successful in detecting \glspl{frb}, but there still exists a strong motivation for using high-gain, single-element telescopes for \gls{frb} surveys. Though the \gls{gbt} has a narrow \gls{fov} compared to that of \gls{askap} or Parkes, the \gls{gbt} provides a significant increase in sensitivity, allowing for the detection of low-fluence \glspl{frb} that would otherwise be missed. These detections would provide an important contribution to future population studies.

Based on previous detections, \glspl{frb} show no preferred sky direction. As such, commensal data acquisition systems provide augmented science output from an observation at the additional cost of running dedicated hardware \citep{fkg+18}. For a fixed \gls{fov}, maximizing the total observing time also maximizes the event detection rate. A dedicated splitter  for the L-band receiver was recently installed on the \gls{gbt}. This allows for full-time observations using this receiver even when other receivers are at the Gregorian focus. These observations at lower, but stil reasonable, sensitivity still provide vital time on the sky, hence maximizing the possibility of an \gls{frb} detection. 

In this work we discuss the design and implementation of the commensal \gls{frb} search back-end using the dedicated L-band tap. We discuss the signal path and processing pipeline in \S \ref{sec:sys}. Initial test observations to verify the pipeline are covered in \S \ref{sec:obs}. We also discuss the survey sensitivity and expected detection rates in \S \ref{sec:sensitivity}.

\section{System Description}
\label{sec:sys}

\GB~is inspired from its predecessor at the Arecibo Telescope, SETIBURST \citep{cmc+17}, and works in parallel with a duplicate SERENDIP VI system at the \gls{gbt}.  Although the concept for \GB~is derived from SETIBURST, it differs in implementation significantly. The back-end consists of three sub-systems, the commensal modifications at the front-end, the ROACH2 \gls{fpga} board\footnote{https://casper.berkeley.edu/wiki/ROACH2} for signal processing, and an \gls{frb} search system. We describe each of these in the subsections below. 

\subsection{L-band Receiver Commensal Modifications}
\label{subsec:tap} 

The \gls{gbt} has an unblocked aperture with an off-axis arm containing an eight-position feed turret. The front-end analogue electronics are situated below a secondary reflector, 8~m in diameter. A circular turret with eight positions is used to select the primary observing receiver. In an earlier test, \cite{m13} found that apart from the focus position, the L-band (1--2~GHz) feed can be used for commensal observing in four more positions with decreasing sensitivity relative to the focus position. We have added a directional coupler to the L-band feed to allow commensal mode observations even when the feed is not being used as the primary observing feed.

\begin{figure*}
\centering
\includegraphics[scale=0.35]{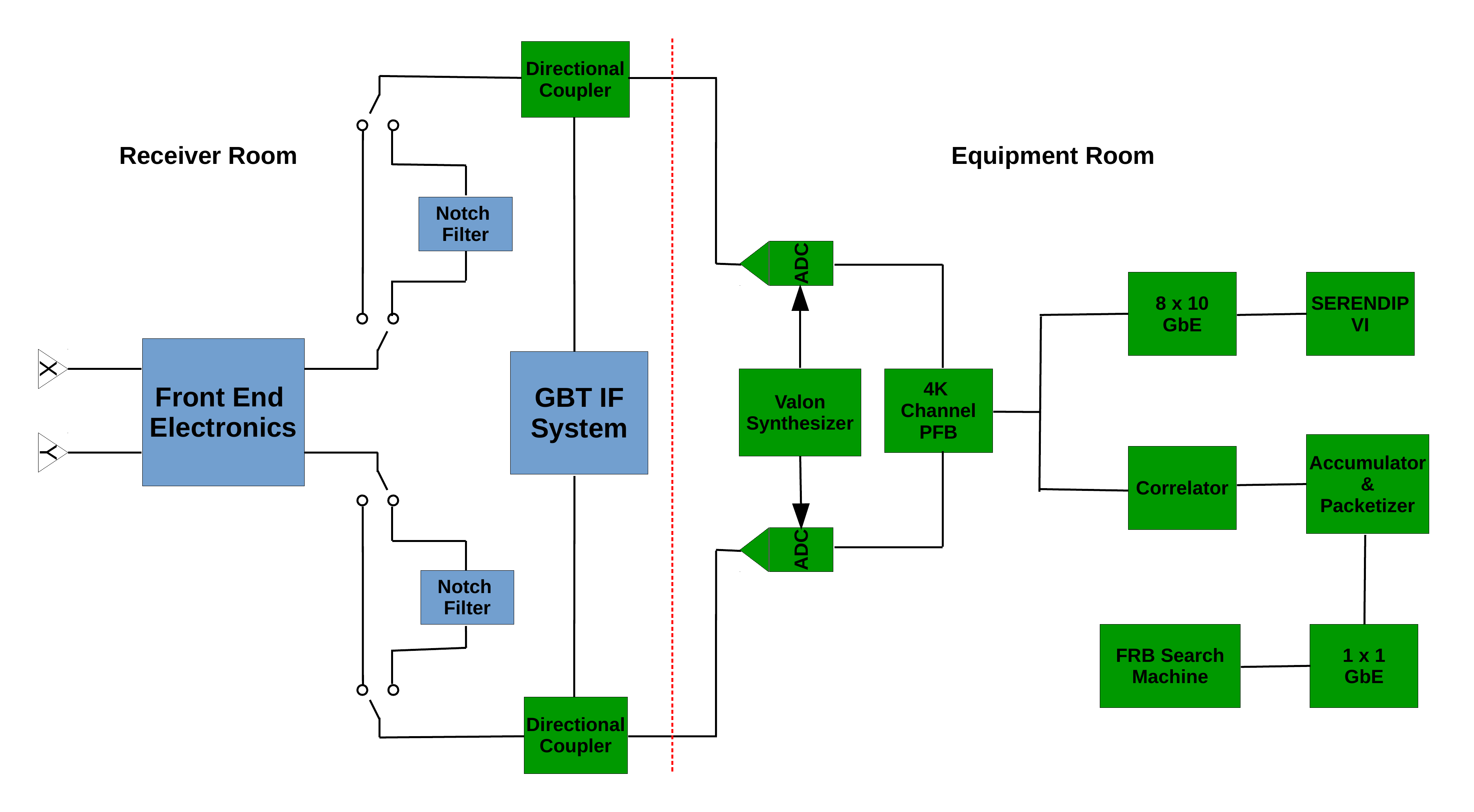}
\caption{Block diagram showing the signal chain for the regular (blue components) and commensal mode (green components) for the L-band receiver. The signal chain on the left of the dashed red line is located in the receiver room, while the chain on the right is located in the equipment room. The signal travels on an optical fiber from the receiver room to the equipment room.}
\label{fig:tap}
\end{figure*}

The dual-polarization L-band \gls{rf} signal received from the feed (denoted as X and Y in Fig.~\ref{fig:tap}) travels through the front-end electronics which consists of an optional circular polarization synthesizer, \gls{lna}, noise calibrator, and a user-selectable notch filter.\footnote{The notch filter is a user-selectable filter to suppress a known radar signal in the frequency band between 1200--1310~MHz}  We installed two directional couplers (one for each of the polarizations) with 20~dB gain in the L-band signal chain between the notch filter  and the \gls{if} system (see Fig.~\ref{fig:tap}).  The output from the coupler is then transmitted to the equipment room through an optical fibre and digitized.  We added isolators before the couplers to minimize reflections which were inducing ripples across the receiver bandpass. 

We then conducted tests to ensure that the additional components do not affect regular observations. In these tests, we observed in the direction of the North celestial pole at night, under good weather. This reduced the possibility of introducing changes in spillover from ground, elevation/atmosphere, the  sky, and baseline shapes from standing waves in the optics due to solar illumination. We conducted an 11-minute observation using the \gls{dcr} to cover the entire L-band frequency range of 1150--1730~MHz. The addition of the coupler resulted in a negligible change in system temperature. We also conducted 40-minute observations in the single window spectral mode at centre frequencies of 1365, 1400, 1420 and 1665~MHz with a bandwidth of 20~MHz to ensure that the spectral-line data quality did not suffer due to the insertion of the couplers. We concluded that the insertion of the couplers in the signal chain did not degrade the data quality.

\subsection{Modified SERENDIP VI FPGA Design}
\label{subsec:fpga}

\begin{figure}
    \centering
    \includegraphics[scale=0.6]{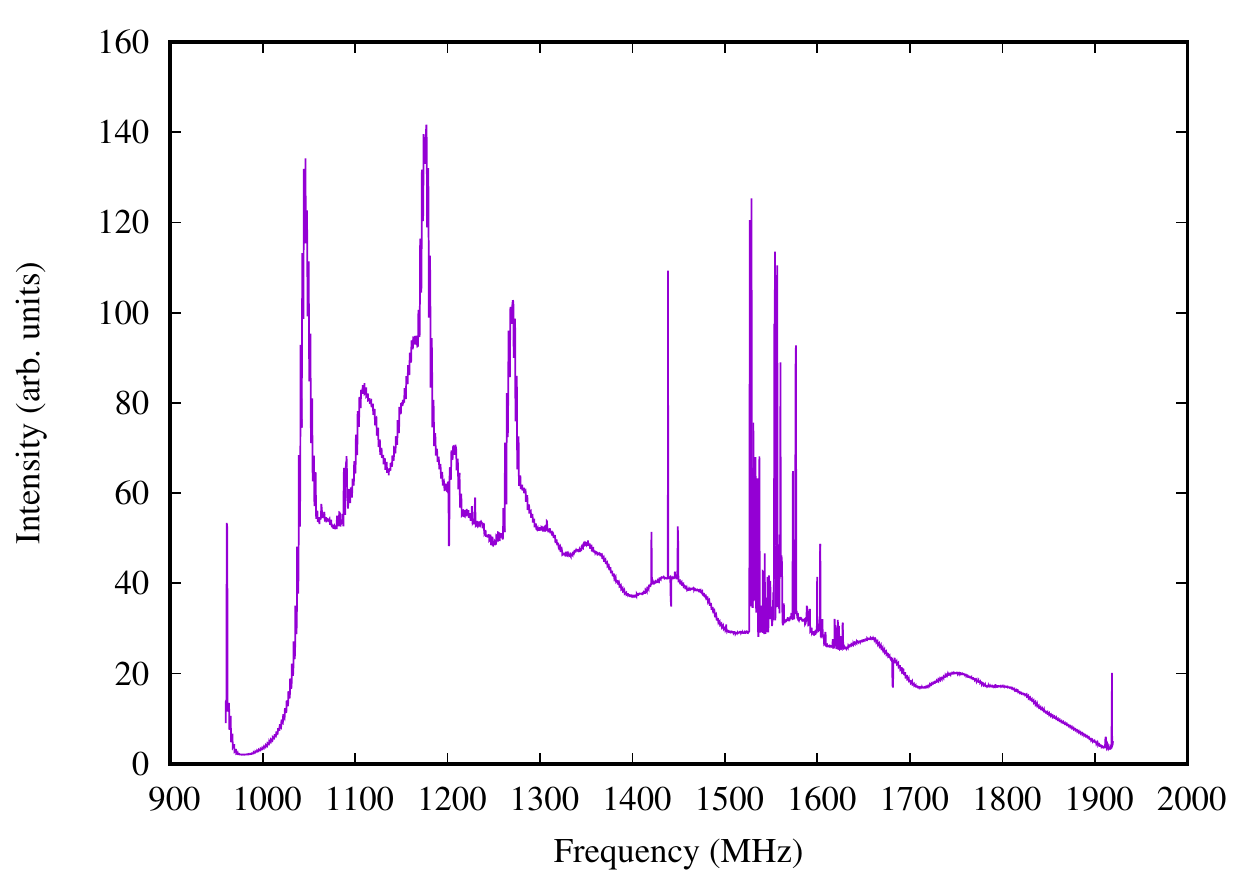}
    \caption{Bandpass response of the \GB~ system from an observation carried out when the L-band feed was at the Gregorian focus. The bandpass shape is similar at other turret positions but with noise statistics proportional to $T_{{\rm sys}}$.}
    \label{fig:bandpass}
\end{figure}

A ROACH2 board is used to perform \gls{dsp} on the L-band signal. This board is primarily used for the SERENDIP VI system \citep[see][for more details]{cmc+17} but we have updated the firmware with additional capabilities for \GB. The analogue signal digitized with an 8-bit \gls{adc} at 1920~Msps. A Valon frequency synthesizer is used as the sampling clock. The second Nyquist zone is used to sample the 960--1920~MHz band. A \gls{pfb} is used to channelize the band into 4096 frequency sub-bands. The sub-bands are then power detected and accumulated to produce a spectrum every 256~$\mu$s (see Fig.~\ref{fig:bandpass} for an example). The spectrum is then packetized and transmitted as \gls{udp} packets over a 1 gigabit Ethernet link to the \gls{frb} search machine. A complete spectrum is divided into 16 packets. Each packet contains 256 spectral channels for each polarization with a total of 768 bytes of data (256 bytes header, 256 bytes power of X and 256 bytes power of Y). This results in a data rate of 384~Mbps. The FPGA design is publicly available\footnote{https://github.com/SparkePei/dibas-upgrade-frb}. 

\subsection{FRB Search Machine}
\label{subsec:srch}

The \gls{frb} search machine is composed of two Intel Xeon E5--2640 processors, an NVIDIA GeForce GTX Titan Xp \gls{gpu} with Pascal architecture and 12~GB memory, a 1.2~TB solid-state drive for temporary storage and a 10~TB additional hard drive for long-term storage.

We use a custom receiver code\footnote{https://github.com/mpsurnis/greenburst} to capture the incoming \gls{udp} packets and reconstruct the spectra. To do this, we use two static buffers, each of length $2^{17}$ samples. This corresponds to 33.6 seconds of time, which is the dispersion delay across our band at a DM of 10000 \pcc. The static buffers are used in an alternate fashion, i.e. at a given time, one buffer is used to save a copy of the incoming spectra while the other is used to write the previous data block to disk. The code writes $2^{21}$ spectra (corresponding to an integration time of $\sim$537~s) to a single filterbank file. In order to avoid losing candidate information in the last data block, we have implemented an overlap of 33.6 seconds between two consecutive filterbank files. Once a filterbank file is created, the search pipeline carries out a single-pulse search covering a \gls{dm} range of 10$-$10000~\pcc. This search is performed using the \gls{gpu}-accelerated \textsc{heimdall} program. The pipeline then sifts through potential candidates and produces plots for most likely \gls{frb} candidates. Further details of the completed pipeline will be discussed in a future paper.

\section{Initial Observations and Sensitivity Tests}
\label{sec:obs}

\begin{figure}
    \includegraphics[width=1.0\linewidth]{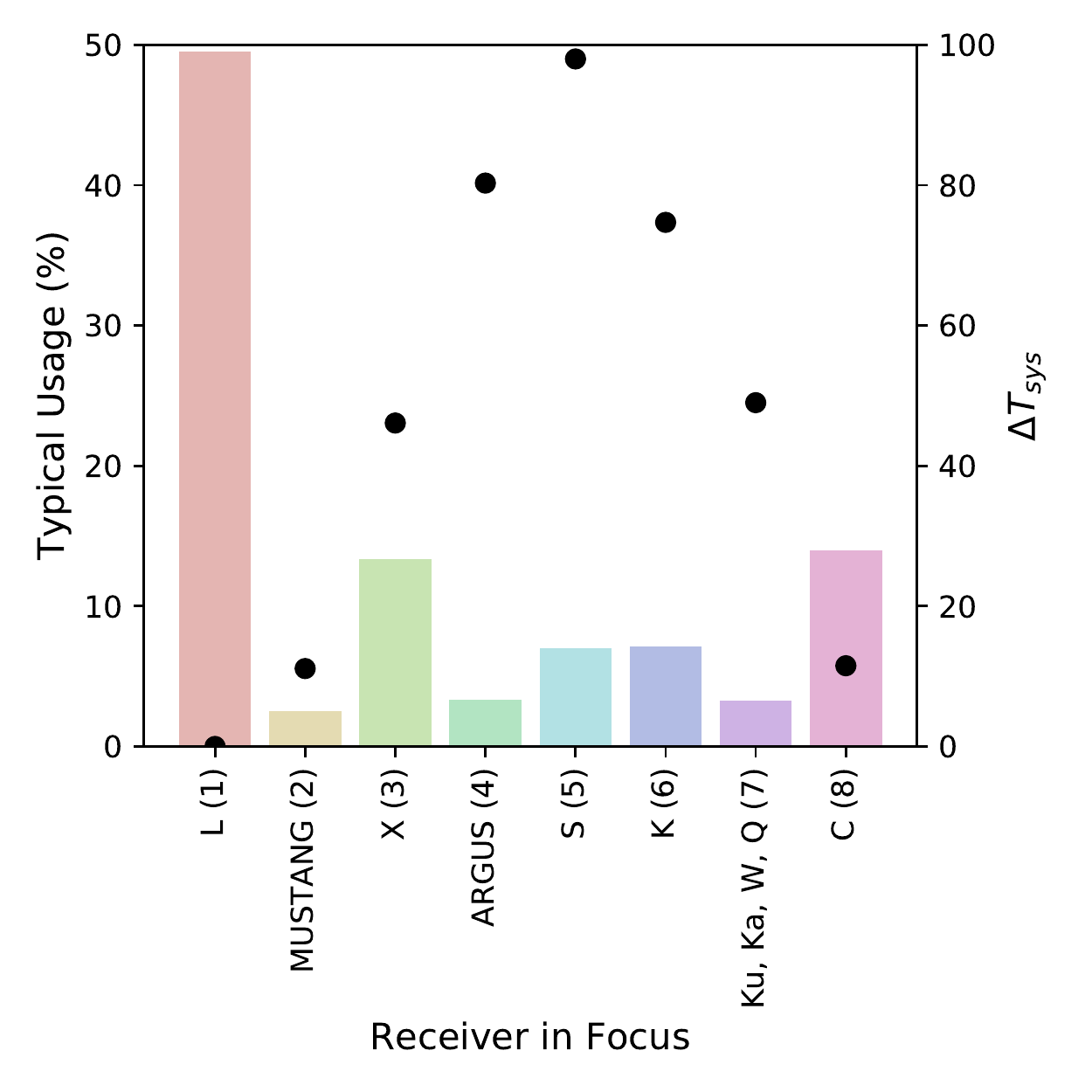}
    \caption{Excess system temperature $\Delta T_{\textrm{sys}}$ as a function of the receiver in focus/turret position for the \gls{gbt} L-band receiver (black points). Expected typical receiver usage based on 2018 usage (bars).}
    \label{fig:turret_usage_sens}
\end{figure}

We measured the sensitivity of \GB~by switching between the eight turret feed positions and observing the same calibrator source. We measured the effective system temperature $T_{{\rm sys}}$ for the L-band receiver using the \gls{dcr} at 1375~MHz with a 20~MHz bandwidth and a sampling time 200~ms to observe the calibrator source 3C147. The additional noise $\Delta T{{\rm sys}}$ due to the turret position offset (relative to the L-band feed at the focus) is plotted as black dots in Fig.~\ref{fig:turret_usage_sens} and listed in Table~\ref{tab:sys_params}. The beam shape of the feed is slightly broadened at the offset turret positions due to axial de-focusing. We measured this for each turret position using the calibrator scans. We list the turret frame offsets and measured \gls{fov} (equivalent width at the half power point) of the L-band feed in Table \ref{tab:sys_params}. Observing with the L-band feed when in neighbouring turret positions (C-band, MUSTANG) results in approximately 30\% decrease in sensitivity. In the worst case, when the S-band feed is at the focus, the effective sensitivity of the L-band system is equivalent to a single dish with a diameter of about 28~m.

\begin{table*}
    \caption{A summary of the relevant parameters for all GBT turret positions. From left to right, the columns list turret position, feeds currently in position, approximate turret rotation angle corresponding to the position offset from the L-band position, the offsets in azimuth and elevation, the excess system temperature for the L-band receiver as compared to the focus position, the measured \gls{fov} of the telescope beam, the antenna temperature, aperture efficiency, telescope gain, estimated sensitivity at the half-power point, usage based on total on-sky time in 2018 and our estimated FRB rate (see text).} 
    \centering
    \begin{tabular*}{\textwidth}{@{}c\x c\x c\x c\x c\x c\x c\x c\x c\x c\x c\x c\x c@{}}
    \hline \hline
    Turret   & Receiver &  Rotation          & $\Delta$ Az & $\Delta$ El & $\Delta T_{{\rm sys}}$  & FoV  & $T_{{\rm A}}$ & $\eta$ & G      & $S_{{\rm min}}$ &  Usage & FRB rate \\
    position & ~        & angle ($^{\circ}$) & (')         & (')         & (K)                     & (')  & (K)           & ~      & (K/Jy) & (Jy)            & (\%)   &  (yr$^{-1}$) \\
    \hline
    1        & L-band  & 0    &  ~0.0       &    ~0.0   &  0.0  &  9.2 & 37.45  & 0.70   & 2.0    &  0.130  & 41.7  & 2$-$6 \\
    2        & MUSTANG & 300  & ~22.5       & $-$13.2   & 11.1  &  9.4 & 28.06  & 0.52   & 1.4    &  0.236  &  2.1  & 0.05$-$0.2 \\
    3        & X-band  & 260  & ~25.3       & $-$30.2   & 46.1  &  9.5 & 14.01  & 0.26   & 0.7    &  0.794  & 11.2  & 0.1$-$0.3 \\
    4        & ARGUS   & 220  & ~16.2       & $-$45.7   & 80.3  & 10.8 &  8.26  & 0.15   & 0.4    &  1.936  &  2.8  & 0.01$-$0.04 \\
    5        & S-band  & 180  & ~0.0        & $-$51.5   & 98.0  & 10.8 &  4.28  & 0.08   & 0.2    &  4.454  &  5.9  & 0.01$-$0.04 \\
    6        & K-band  & 140  & $-$16.9     & $-$45.2   & 74.7  & 10.3 &  6.97  & 0.13   & 0.4    &  1.856  &  6.0  & 0.02$-$0.08 \\
    7        & Ku-band$^{{\rm \dagger}}$ & 100 & $-$25.4    & $-$29.8   & 49.0  &  9.6 & 14.94   & 0.28   & 0.8    &  0.718   &  2.7   & 0.02$-$0.08 \\
    8        & C-band &  60   & $-$22.1     & $-$12.5   & 11.5 &  9.3 & 28.84   & 0.54   & 1.5    &  0.224  & 11.8  & 0.3$-$1.0 \\
    \hline \hline
    \end{tabular*}
    \label{tab:sys_params}
    \medskip
    \tabnote{$^{{\rm \dagger}}$ Turret position 7 also houses Ka-band, W-band and Q-band feeds in rotation.}
\end{table*}

Fig.~\ref{fig:turret_usage_sens} also shows the typical percentage usage of the feeds based on 2018 usage statistics (T.~Minter, private communication). Turret positions 4 (ARGUS), 5 (S-band), and 6 (K-band) are partially blocked by the primary focus structure resulting is significantly reduced sensitivity. But, only approximately 15\% of the 2018 observing time was used for observations with these feeds. Nearly half of the observing time was using the L-band feed, as such most \GB~observations will be at the optimal sensitivity. 

\begin{center}
\begin{figure}
    \includegraphics[width=1.0\linewidth]{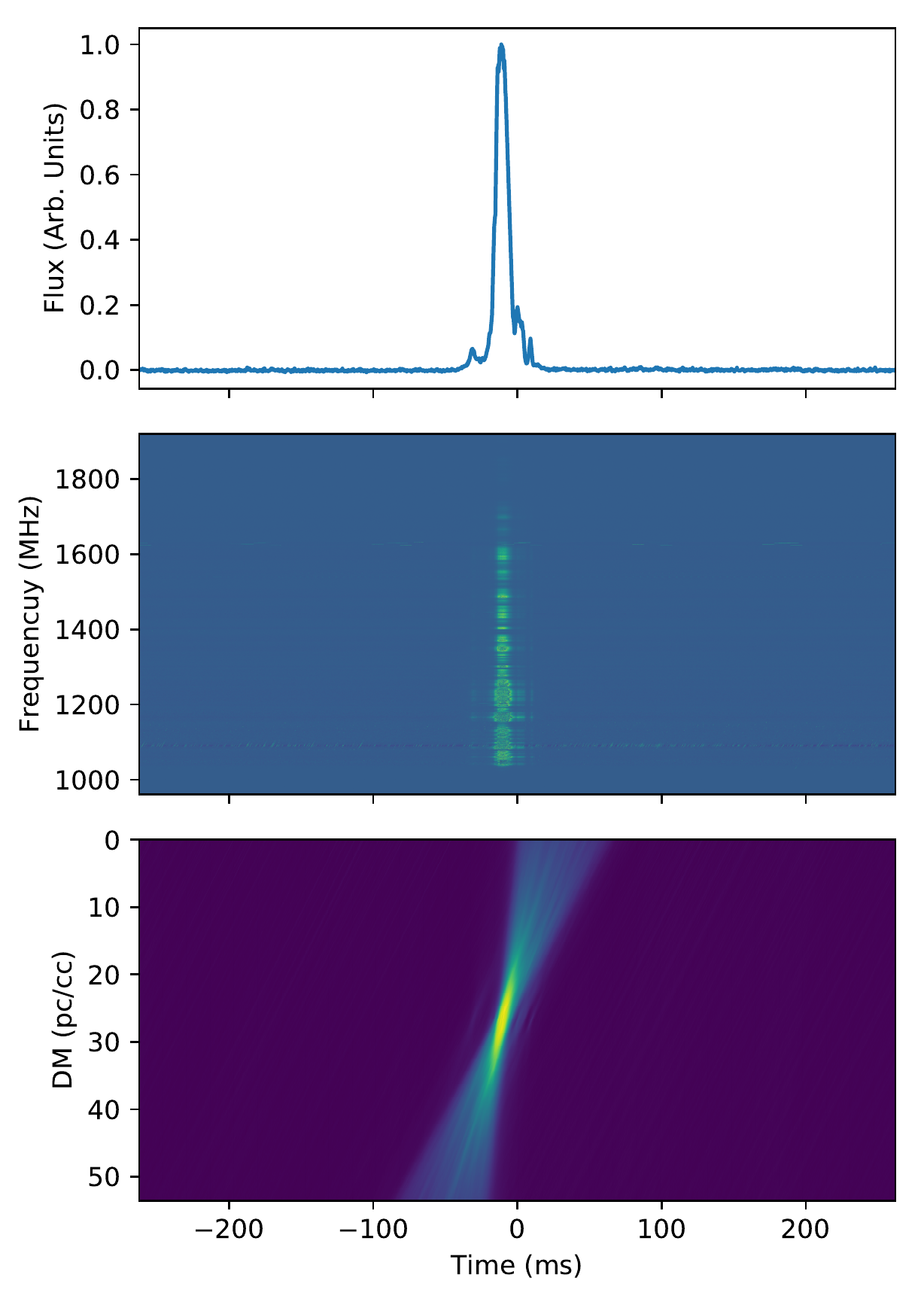}
    \caption{Candidate plot showing a single pulse detected from PSR\,B0329+54 with the \GB~ pipeline. Top panel shows the pulse in time, middle panel shows the de-dispersed pulse as a function of frequency, and the bottom panel shows the pulse in DM-time plane.}
    \label{fig:0329}
\end{figure}
\end{center}

To test the \gls{frb} search pipeline, we observed PSR\,B0329+54. An example candidate plot from the real-time single pulse search pipeline is shown in Figure \ref{fig:0329}. Our single-pulse search pipeline will produce similar plots for all the potential \gls{frb} candidates. 

\section{Event Rate Predictions}
\label{sec:sensitivity}

\begin{center}
\begin{figure}
    \includegraphics[scale=0.65]{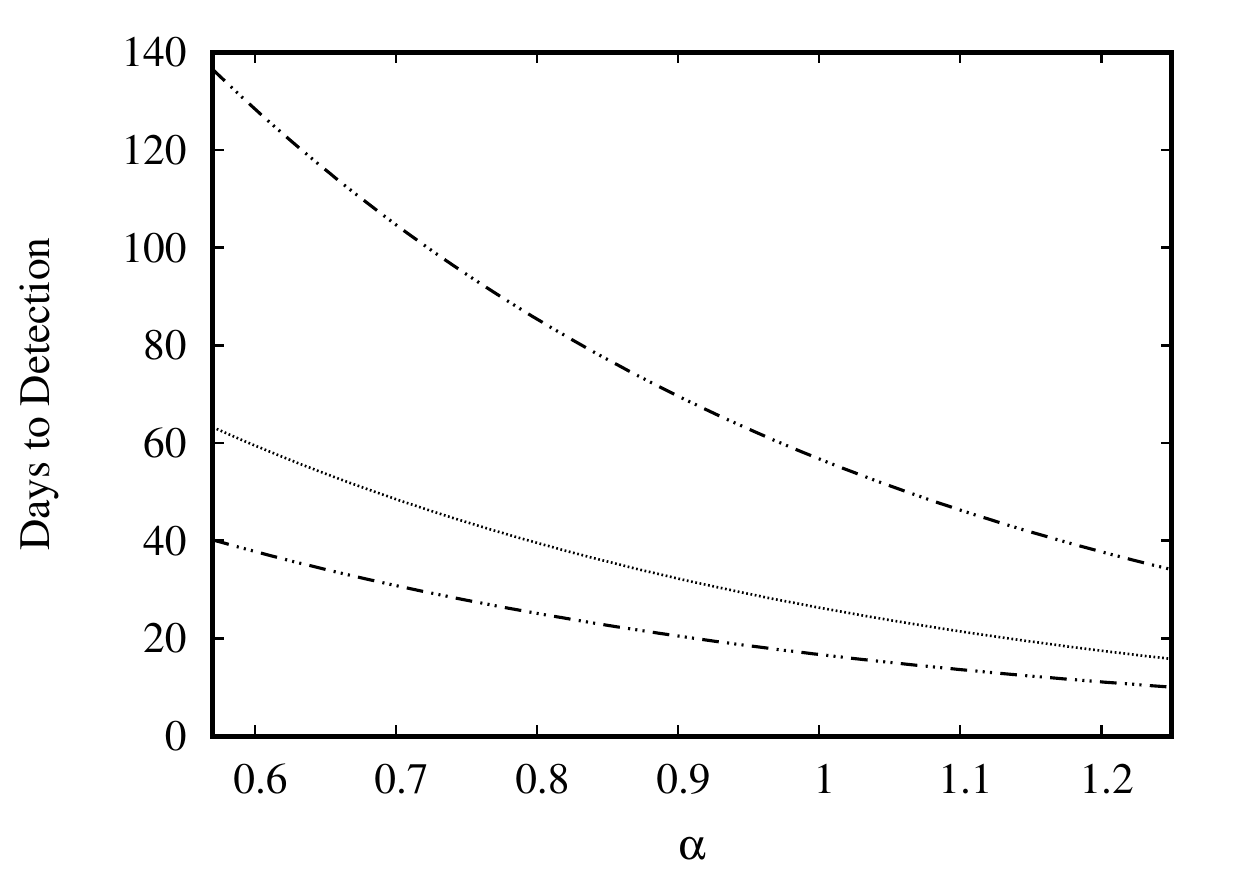}
    \caption{Predicted number of days to the first \gls{frb} detection with \GB~as a function of $\alpha$. The dotted line in the middle denotes the mean while the dot-dashed lines denote the 95\% confidence intervals}
    \label{fig:alpha}
\end{figure}
\end{center}

To estimate the expected \gls{frb} rate with \GB, we first use the radiometer noise considerations
to compute the minimum detectable flux density
\begin{equation}
S_{{\rm min}}  =  \frac{S/N_{{\rm min}} T_{{\rm sys}}}{G \sqrt{n_{p}t_{{\rm obs}}\triangle f}},
\label{eqn:radiometer}
\end{equation}
where we assume a system temperature ($T_{{\rm sys}}$) of 40~K 
(including a cold sky temperature of 10~K), a cut-off signal to noise ratio (S/N) of 10 ($S/N_{{\rm min}}$), telescope gain ($G$) of 2 K~Jy\sups{$-$1}, with 2 polarizations summed ($n_p=2$) over a bandwidth ($\triangle f$) of 960 MHz and a typical \gls{frb} pulse width ($t_{{\rm obs}}$) of 5 ms. This results in an $S_{{\rm min}}$ of 65 mJy at the bore-sight or 130 mJy at the half power point of the beam. Re-arranging Equation 6 from \cite{lvl+17}, replacing $a$ by $R_{0}$ and $b$ by $\alpha$, above a flux density $S$, we get the expected \gls{frb} rate
\begin{equation}
R(> S) = \frac{\pi r^2}{\alpha \ln(2)} R_{0} \left(\frac{S}{1~\rm{Jy}}\right)^{-\alpha},
\end{equation}
where $R_{0}$ is the all sky rate of \glspl{frb} and $\alpha$ is the (currently uncertain) slope of the source count distribution. For a population of sources uniformly distributed in Euclidean space, $\alpha=1.5$. We have also assumed a circular beam with a Gaussian power pattern with $r$ being the half width at half maximum. Adopting a value of 4.6$^\prime$ for $r$ (see Table \ref{tab:sys_params}), 0.91 for $\alpha$ and a maximum sky rate $R_0$ of 924 \gls{frb} events per sky per day \citep{lvl+17} results in a maximum rate of 5 $\times$ 10\sups{$-$2} \gls{frb} events per day for \GB. Inverting the rate gives us the minimum wait time of 20 days before the first \gls{frb} detection. Figure \ref{fig:alpha} shows the ranges of the wait times based on the 95\% confidence intervals on $\alpha$ and $R_0$ as reported by \cite{lvl+17}. For a mean value of 0.91 for $\alpha$ and 587 \gls{frb} events per sky per day for $R_0$ \citep[as reported by][]{lvl+17}, we get a mean wait time of 32 days of continuous operation at the Gregorian focus (turret position 1) for \GB. Using the method described above, we have estimated $S_{{\rm min}}$ and the projected \gls{frb} rates for other turret positions in Table \ref{tab:sys_params}. We expect a combined expected \gls{frb} rate of 2--7 per year of operation.

\section{Discussion}
\label{sec:disc}

We have estimated the wait time for the first \gls{frb} detection adopting  the all-sky rate from \cite{lvl+17}. Based on the usage statistics for the \gls{gbt} in year 2018, the L-band feed was used for scientific observations for approximately 2970 hours out of a total of about 7125 hours of on-sky time (T.~Minter, private communication). For the 95\% confidence interval range of the all-sky \gls{frb} rates \citep{lvl+17}, we project that \GB~ would detect between 2$-$6 \glspl{frb} per year. This rate estimate includes the time L-band was used at the Gregorian focus. The adjacent turret positions (position 2 and 8) constitute an additional 1000 hours of observing time. With an 11~K excess $\Delta T_{\rm{sys}}$, this would result in one additional \gls{frb} per year. The remaining time at other turret positions may result in more serendipitous detections, especially of low-DM and high-brightness events such as the recently discovered \gls{askap} \citep{smb+18} and CHIME \citep{abb+19a} \glspl{frb}. Given this rate, even a non-detection would be useful in constraining the value of $\alpha$ towards the higher end in Figure \ref{fig:alpha}. Assuming that 2018 usage indicates typical GBT usage, we get about 123 days of observations at the maximum sensitivity. Even if half of these data are corrupted by \gls{rfi}, a non-detection would exclude all values of $\alpha > $ 1. A limited number of detections in the first year would also help putting useful constraints on the values of $R_0$, which is the instrument-independent all-sky \gls{frb} rate. Thus, \GB~ is a promising \gls{frb} instrument regardless of the outcome.

Most current \gls{frb} back-ends search multiple beams (e.g.~\gls{askap}, \gls{chime}, Parkes, ALFABURST etc.), while \GB~is a single-beam survey. Although this reduces sky coverage, the high sensitivity of the \gls{gbt} still means it has the potential to detect \glspl{frb}. This increases the search volume in redshift and thus compensates for the lack of sky coverage. If a repeating \gls{frb} were detected with \GB, the narrow telescope beam would reduce the prospective sky area for follow-up observations. This would speed-up the identification of the host galaxy associated with the \gls{frb}. 

The \GB~ pipeline will carry out a blind search for single-pulses over a large \gls{dm} range. This range includes Galactic \glspl{dm} as well. Thus, in addition to \glspl{frb}, our pipeline may also discover radio transients in our Galaxy. The distinct advantage provided by the commensal mode observing is that \GB~ would be recording data all the time, enabling it to make serendipitous discovery of a Galactic radio transient. We have already made such a discovery with the ALFABURST while the telescope was slewing between targets \citep{fkg+18}. In addition, \GB~ may detect more radio pulses from known rotating radio transients (RRATs)\footnote{http://astro.phys.wvu.edu/rratalog}. 

We have a working pipeline, and at the time of writing, have acquired data for a week. We are currently carrying out data quality checks and tuning our search pipeline to the observed \gls{rfi} in the data. We are also ascertaining whether all the metadata are properly logged by cross-referencing them with the GBT logs. \GB~ is exiting its commissioning phase and would soon start searching for \glspl{frb} in real-time.


\begin{acknowledgements}
We thank West Virginia University for its financial support of GBT operations, which enabled some of the observations for this project. M.P.S., M.A.M. and D.R.L. acknowledge support from NSF RII Track I award number OIA$-$1458952. M.P.S., M.A.M., G.G. and D.R.L. are members of the NANOGrav Physics Frontiers Center which is supported by NSF award 1430284. Berkeley efforts were supported by NSF grants 1407804 and 1711254, as well as the Marilyn and Watson Alberts SETI Chair funds. The Green Bank Observatory is a facility of the National Science Foundation operated under cooperative agreement by Associated Universities, Inc.
\end{acknowledgements}

\bibliographystyle{pasa-mnras}
\bibliography{ref}

\end{document}